\documentclass[namedreferences]{solarphysics}

\usepackage[hyperref,optionalrh,showbiblabels]{spr-sola-addons} % For Solar Physics 

\usepackage{amsmath}

\usepackage{graphicx}        % For eps figures, newer & more powerfull
\usepackage{color}           % For color text: \color command
\usepackage{breakurl}        % For breaking URLs easily trough lines
\usepackage{multirow}
\usepackage{array}
\usepackage[utf8]{inputenc}
\usepackage{hyperref}

\newcommand{\etal}{{\it et al.}}
\newcommand{\rot}{\rotatebox{90}}

\newcommand{\qq}[1]{``#1"}

\newcommand{\mn}[1]{\text{#1}}

\newcommand{\aap}{    {\it Astron. Astrophys.}}

\newcommand{\apj}{    {\it Astrophys. J.}}

\newcommand{\mnras}{  {\it Mon. Not. Roy. Astron. Soc.}}

\newcommand{\pasj}{   {\it Pub. Astron. Soc. Japan}}

\newcommand{\solphys}{{\it Solar Phys.}}

\newcommand{\atao}{   {\it Ann. Tokyo Astron. Obs.}} 
\newcommand{\ban}{    {\it Bull. Astron. Inst. Neth.}}
\chardef\us=`\_

\begin{document}
	
\begin{article}
\begin{opening}

\title{Two Practical Methods for Coronal Intensity Determination}

\author[addressref={aff1},corref,email={hcakmak@istanbul.edu.tr}]{\inits{H.}\fnm{Hikmet}~\lnm{Cakmak}\orcid{0000-0002-1959-6049}}

%\author{\inits{}\fnm{}~\lnm{}\orcid{}}
%\author{Hikmet~\surname{\c{C}akmak}$^{1}$}

%\institute{$^{1}$ Department of Astronomy and Space Science, Istanbul University Faculty of Science
%				34116 Beyaz{\i}t / Istanbul - Turkey\\
%                email: \url{hcakmak@istanbul.edu.tr} 
%              $^{2}$ Second affiliation
%                     email: \url{e.mail-c} \\
%            }
\address[id=aff1]{Department of Astronomy and Space Science, Faculty of Science,  
Istanbul University, Beyaz{\i}t / Istanbul - Turkey}

\runningauthor{H. \c{C}akmak}
\runningtitle{Two Practical Methods for Coronal Intensity Determination}

\begin{abstract}
Determining the relative brightness of the solar corona is one of the most critical stages in solar eclipse studies. For this purpose, images taken with different exposures and polarization angles in white-light observations are used. The composite image of each polarization angle is produced by combining the images of different exposures. With the help of the intensity calibration function of these images, the relative intensity of the corona can be calculated. The total brightness of the solar corona is calculated using Stokes parameters obtained from intensity values of three polarization angles. In this study, two methods are presented: the first is used to obtain the intensity calibration function of the photographic material using calibration images, and the second is used to calculate the combined intensity values of images taken with different polarization angles. 

\end{abstract}

\end{opening}
%-------------------------------------------------

\section{Introduction}\label{intro}
Total solar eclipses are important celestial events that occur rarely. Their main significance originates from the visibility of the solar corona with the naked eye. The solar corona is the outer and enigmatic part of the Sun's atmosphere. Although its particle density is very low, its temperature is very high and reaches a few million degrees \citep{GAH1976,FAL1993}. In addition, this layer has a brightness of about a millionth of the solar photosphere, which makes it hard to observe \citep{GP2010}. These obscurities render every total solar eclipse significant. The general shape of the solar corona mainly depends on the magnetic field of the Sun. Therefore, variations in the magnetic field distribution around the solar disk cause changes in the observed shape of the corona. These variations are seen in solar eclipse observations as changes in coronal shape from one eclipse to another, depending on the solar magnetic activity or solar cycle. Sometimes many coronal structures are relatively concentrated near the equator during solar minimum, while the coronal structures during maximum are distributed across the whole solar disk. Examples of these cases are shown in Figure 1.   

\begin{figure}[h!]\label{fig:1}
	\includegraphics[width=0.495\textwidth]{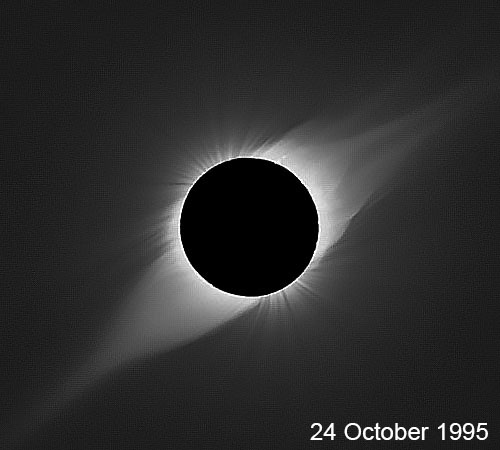}
	\includegraphics[width=0.495\textwidth]{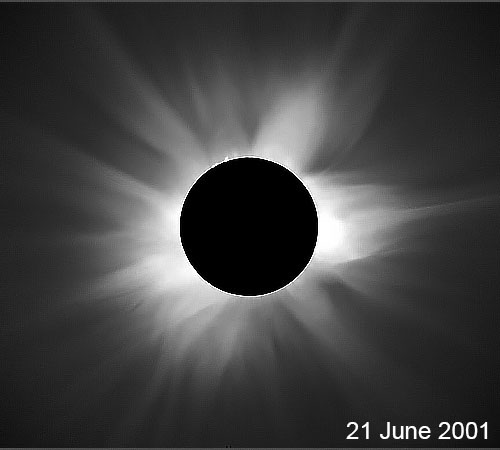}
	\caption{Two example images for the observed shapes of the solar corona during totality. $\it Left$: Minimum. $\it Right$: Maximum phase of the solar cycle. The date on the picture indicates the day of the eclipse.}
\end{figure}

The most prominent feature of the corona is the brightness gradient. In coronal structures smooth transitions in brightness are observed. This is due to the radial distribution of the particles --- mostly electrons --- that disperse above the photosphere. In eclipse studies, the light coming from the corona is accepted as photospheric light scattered to the line of sight by free electrons \citep{VDH1950}. The magnitude of the observed brightness is proportional to particle density \citep{VDH1950,SK1970}. The coronal light brightness determination does not only provide the brightness intensity, but also gives an opportunity to calculate the particle density.

Coronal roll film image brightness determination is a demanding process and has several stages such as bathing and drying films, scanning and importing the data into a digital environment, cleaning and processing the transferred pictures, and finally calibrating the brightness. Most of these stages like the bathing, scanning, and cleaning, are no longer carried out since digital cameras have taken up the role of film cameras in eclipse observations. Todays, the only important stage for the eclipse images is correctly calibrating their brightness. The calibration process is mostly performed by the combined use of both step-wedge densities and exposure-intensity profiles of the films. 

The calibration process depends on predetermined values provided by the film manufacturers. Using the first alternative method brightness calibration calculations are accomplished in a straightforward way, which eliminates the need for the aforementioned values. In this approach, filtered solar disk images taken with different exposures and diaphragm openings are used to bring out the intensity calibration function. The relative brightness of every eclipse image is calculated by using this function of the film. This is explained in Section 2. The second developed method serves the purpose of calculating the composite intensity values of each polarization angle. All appropriate information on this subject is given in Section 3. Finally, total brightness values of the observed coronal light are obtained {\it via} Stokes parameters described for polarized electromagnetic radiation. Necessary explanations on this subject are given in Section 4, along with the formulas used to obtain the total brightness, polarization degree and polarization angle values.

\section{Alternative Method for Intensity Calibration}\label{sec:2}
Although not as common as in the past, classic black-and-white roll films are used in some eclipse observations, and 35 mm or 6$\times$6 cm roll films of Kodak or Ilford are preferred for this purpose. Roll films are made of light-sensitive material, and this sensitivity shows variations depending on observational conditions. This is crucial for eclipse observations, because very many images are taken with different exposures: before, after and during totality. The weather conditions, temperature of the observation location and warming of the equipment change over time. All of these effects cause variances in the sensitivity of photographic material during the observation.

Different exposure times are used to reveal whole coronal structures as much as possible. Short exposures are used for bright parts of the corona and long exposures for the faintest parts. In the solar disk shots that are taken for calibration, however, the saturation effect in long exposures is more prominent than in eclipse images. To reduce this effect, many diaphragms with different apertures are placed in front of the telescope. The number of diaphragm openings is adjusted according to the telescope aperture. For large areas more diaphragms are used, and for small areas fewer are used. When these are identified,  filtered solar disk images are taken with predetermined exposures for every selected diaphragm opening. Some pictures taken at the total solar eclipse of 29 March 2006 (hereafter referred to as TSE2006) are shown in Figure 2. 
\begin{figure}[t!]\label{fig:2}
	\centering
	\includegraphics[width=0.93\textwidth]{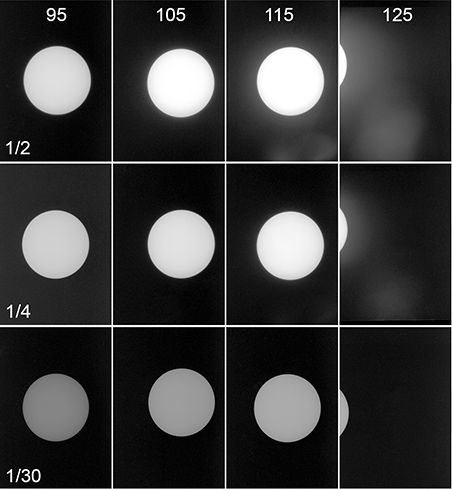}
	\caption{Some samples of solar disk images taken at TSE2006. Numbers on the left are exposure times in seconds, and numbers at the top are the radius of diaphragm openings in millimeters. In the right-most images, the solar disks are shifted to prevent the background from becoming saturated.}
\end{figure}

TSE2006 observation was carried out by the staff of the Astronomy and Space Science Department of Istanbul University in the Manavgat district of Antalya. For this observation, a 35mm Kodak T-max 100 black-and-white roll film in ISO 100/21$^\circ$ format was used. Kodak Professional D-76 Developer was used as the development solution in the bathing process. The film remained in the developer for 4 minutes at room temperature (20-24 $^\circ$C) and was left in the stabilizing solution for approximately 15 minutes after 35-40 seconds in the fixing bath. The last bathing of the film was made, and it was left to dry. All photos were transferred into the computer environment {\it via} the Mikrotek ArtixScan 4000t Film Scanner.

\subsection{Descriptions of Normalized and Relative Intensities}\label{sec:2-1}
The coronal brightness is usually expressed in terms of the observed brightness of the photospheric solar disk and this shows a ratio that indicates the relative intensity. With a similar approach, a new ratio is defined as a \qq{normalized intensity} which is the value of the image intensity divided by its background intensity. Because very many images taken at different exposures and diaphragm openings, however, the background intensity of every image is different. Therefore, a misleading situation would arise if every image were normalized within itself. The intensity changes between different exposures would be neglected. To indicate this situation correctly, the shortest exposure background intensity value is selected as a unit background intensity. Thus, the normalized intensity ($I_\mn{N}$) is defined by
\begin{equation} \label{eq:1}
	I_\mn{N}=\frac{I_\mn{0}}{I_\mn{min}} \, ,
\end{equation}
where $I_0$ is the average intensity value of the considered area and $I_\mn{min}$ is the average intensity value of the shortest exposure background. 
\begin{figure}[b]\label{fig:3}
	\centering
	\includegraphics[width=0.35\textwidth]{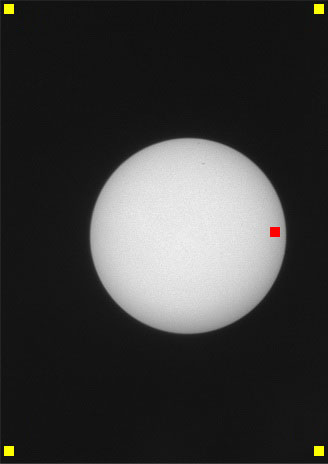} \quad
	\includegraphics[width=0.35\textwidth]{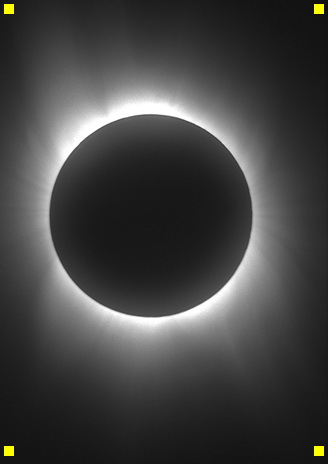}
	\caption{Examples of selected areas to calculate average intensity values for a solar disk image ($\it left$) and an eclipse image ($\it right$). Yellow boxes are indicating the areas in the background and red box is the area on the disk.}
\end{figure}

The average intensity of a solar disk image is calculated as the mean value of the pixel intensities within a particular area on the disk. This particular area can be any desired width such as 50 $\times$ 50 or 100 $\times$ 100 pixels in size, but it must be large enough to represent a general average value. The location of the area on the disk must be chosen carefully: it must not be close to the center or the edge of the disk because of saturation effects in the center and because of the limb-darkening effect at the edge. To calculate the average background intensity value, four different areas are selected that lie fairly close to the four corners of the disk frame, and the average value of these areas is taken as an average background value of the frame. Samples of such areas are shown in Figure 3 as yellow (corners) and red (disk) boxes.

The brightness measured in the focal plane of a telescope depends on various parameters such as sensitivity of the used photographic material, exposure time, and light transmissions of the solar filter and polarizer. The instrument aperture and diaphragms and the average observed solar disk intensity should also be taken into account. With the aid of the explanations given in the books by \cite{MWR1994}, the \qq{Introduction to Radiometry and Photometry}, and by \cite{AMK2007} the \qq{Field Guide to Illumination}, the observed solar disk intensity $I_\mn{0}$ can be defined by
\begin{equation} \label{eq:2}
	I_0 \approx \bar{I}_{\odot} \, f_\mn{int} \, f_\mn{pol} \, \left( \frac{A_\mn{D}}{A_0} \right) \, t \, ,
\end{equation}
where $\bar{I}_{\odot}$ is average intensity of the apparent solar disk, $f_\mn{int}$  and $f_\mn{pol}$ are the light transmissions of the solar filter and polarizer, respectively, $\it t$ is the exposure time, and $A_0$ and $A_\mn{D}$ are the area of telescope aperture and diaphragm openings, respectively. 

In accordance with Equation 2, the relative intensity $I_\mn{R}$ is defined as the ratio of the observed disk intensity to the average intensity of the apparent solar disk, and it is given by
\begin{equation} \label{eq:3}
	I_\mn{R}=\frac{I_\mn{0}}{\bar{I}_{\odot}} \approx f_\mn{int} \, f_\mn{pol} \, \left( \frac{A_\mn{D}}{A_0} \right) \, t \, .
\end{equation}
Here, the relative intensity depends on the exposure time and materials used in the observation. From a general point of view, relative and normalized intensities show the same brightness from different perspectives. In other words, the average intensity of an energy source is shown by the relative intensity from an analytical aspect and the normalized intensity from an observational aspect. 

When a graph is drawn between the relative and normalized intensities, an intensity calibration function is obtained by fitting a least-squares curve. With this function, it is possible to convert any normalized intensity into relative intensity. At this point, giving some numerical samples is more appropriate to facilitate understanding of the relationship between them.

\subsection{An Example for Normalized and Relative Intensities}\label{sec:2-2}
The measured intensity values of the solar disk and its background in calibration images taken at TSE2006 are listed in Table 1. All intensity values are given in pixel intensity which ranges from 0 to 65535. The average background intensity value is measured as 5874. This is the median background value of the shortest exposures (1/125 s).

Normalized intensity values were calculated using Equation 1, and the results are listed in Table 2 with the exposures we used. The relative intensity values were also calculated by considering Equation 3 where $f_\mn{int}$ = 1.6 $\times 10^{-5}$ , $f_\mn{pol}$ = 0.3 and $r_\mn{0}$ = 101.6 mm (telescope aperture radius). Figure 4 was obtained using these normalized and relative intensities. In the graph, every exposure value is shown with different symbols to indicate their locations, and the fit curve is represented by the solid line.

\renewcommand{\arraystretch}{1.02}%
\setlength{\tabcolsep}{12.5pt} 
\begin{table}[htbp]\label{tab:1}%
	%	\centering
	\caption{Measured pixel intensity values of the solar disk and its background in calibration images.}
	\begin{tabular}{p{0.2cm}lrrrrr}
		& Exposure & \multicolumn{5}{c}{Radius of diaphragm opening (mm)} \\
		\cline{3-7}
		& \multicolumn{1}{c}{(s)}  & \multicolumn{1}{c}{95} & \multicolumn{1}{c}{105} & \multicolumn{1}{c}{115} & \multicolumn{1}{c}{125} & \multicolumn{1}{c}{135} \\
		\cline{2-7}
		\multicolumn{1}{c}{\multirow{5}[2]{*}{\rot{Disk}}} & \hspace*{0.2cm} 1/2 & 39204 & 57559 & 61437 & 63859 & 64362 \\
		\multicolumn{1}{c}{} & \hspace*{0.2cm} 1/4   &  3534 & 37385 & 47767 & 54585 & 54712 \\
		\multicolumn{1}{c}{} & \hspace*{0.2cm} 1/30  & 15130 & 17732 & 24757 & 28396 & 28779 \\
		\multicolumn{1}{c}{} & \hspace*{0.2cm} 1/60  & 10888 & 14524 & 19473 & 22617 & 23476 \\
		\multicolumn{1}{c}{} & \hspace*{0.2cm} 1/125 &  7745 & 12321 & 14651 & 17525 & 18082 \\
		\cline{2-7}
		& & & & & &  \\[-0.35cm]
		\cline{2-7}
		\multicolumn{1}{c}{\multirow{5}[2]{*}{\rot{Background}}} & \hspace*{0.2cm} 1/2 & 5812 & 5903 & 6765 & 6689 & 8363 \\
		\multicolumn{1}{c}{} & \hspace*{0.2cm} 1/4   & 6860 & 5740 & 5940 & 6084 & 6842 \\
		\multicolumn{1}{c}{} & \hspace*{0.2cm} 1/30  & 5746 & 5691 & 5703 & 6164 & 6083 \\
		\multicolumn{1}{c}{} & \hspace*{0.2cm} 1/60  & 5969 & 5711 & 5907 & 6214 & 5899 \\
		\multicolumn{1}{c}{} & \hspace*{0.2cm} 1/125 & 5939 & 5874 & 5874 & 6129 & 5742 \\
		\cline{2-7}
	\end{tabular}%
\end{table}%

\setlength{\tabcolsep}{14pt} 
\begin{table}[htbp]\label{tab:2}%
	%	\centering
	\caption{Calculated normalized and relative intensity values of the calibration images.}
	\begin{tabular}{p{0.6cm}lrrrrr}
		& Exposure & \multicolumn{5}{c}{Radius of diaphragm opening (mm)} \\
		\cline{3-7}
		& \multicolumn{1}{c}{(s)}  & \multicolumn{1}{c}{95} & \multicolumn{1}{c}{105} & \multicolumn{1}{c}{115} & \multicolumn{1}{c}{125} & \multicolumn{1}{c}{135} \\
		\cline{2-7}
		\multicolumn{1}{c}{\multirow{5}[2]{*}{\rot{\parbox[t][0.04\textwidth][t]{0.18\textwidth}
					{\centering Normalized Intensity}
		}}} 
		& \hspace*{0.2cm} 1/2 & 6.7 & 9.8 & 10.5 & 10.9 & 11.0 \\
		\multicolumn{1}{c}{} & \hspace*{0.2cm} 1/4 & 6.0 & 6.4 & 8.1 & 9.3 & 9.3 \\
		\multicolumn{1}{c}{} & \hspace*{0.2cm} 1/30 & 2.6 & 3.0 & 4.2 & 4.8 & 4.9 \\
		\multicolumn{1}{c}{} & \hspace*{0.2cm} 1/60 & 1.8 & 2.5 & 3.3 & 3.9 & 4.0 \\
		\multicolumn{1}{c}{} & \hspace*{0.2cm} 1/125 & 1.3 & 2.1 & 2.5 & 3.0 & 3.1 \\
		\cline{2-7}
		& & & & & &  \\ [-0.35cm]
		\cline{2-7}
		\multicolumn{1}{c}{
			\multirow{5}[3]{*}{\rot{\parbox[t][0.04\textwidth][t]{0.18\textwidth}
					{\centering Relative Int. ($\times 10^{-9} \bar{I}_{\odot}$) }
		}}} 
		& \hspace*{0.2cm} 1/2 & 526 & 642 & 770 & 910 & 1061 \\
		\multicolumn{1}{c}{} & \hspace*{0.2cm} 1/4 & 263 & 321 & 385 & 455 & 531 \\
		\multicolumn{1}{c}{} & \hspace*{0.2cm} 1/30 & 35 & 43 & 51 & 61 & 71 \\
		\multicolumn{1}{c}{} & \hspace*{0.2cm} 1/60 & 18 & 21 & 26 & 30 & 35 \\
		\multicolumn{1}{c}{} & \hspace*{0.2cm} 1/125 & 8 & 10 & 12 & 15 & 17 \\
		\cline{2-7}
	\end{tabular}%
\end{table}%

\begin{figure}[b!]\label{fig:4}
	\centering \includegraphics[width=0.67\textwidth]{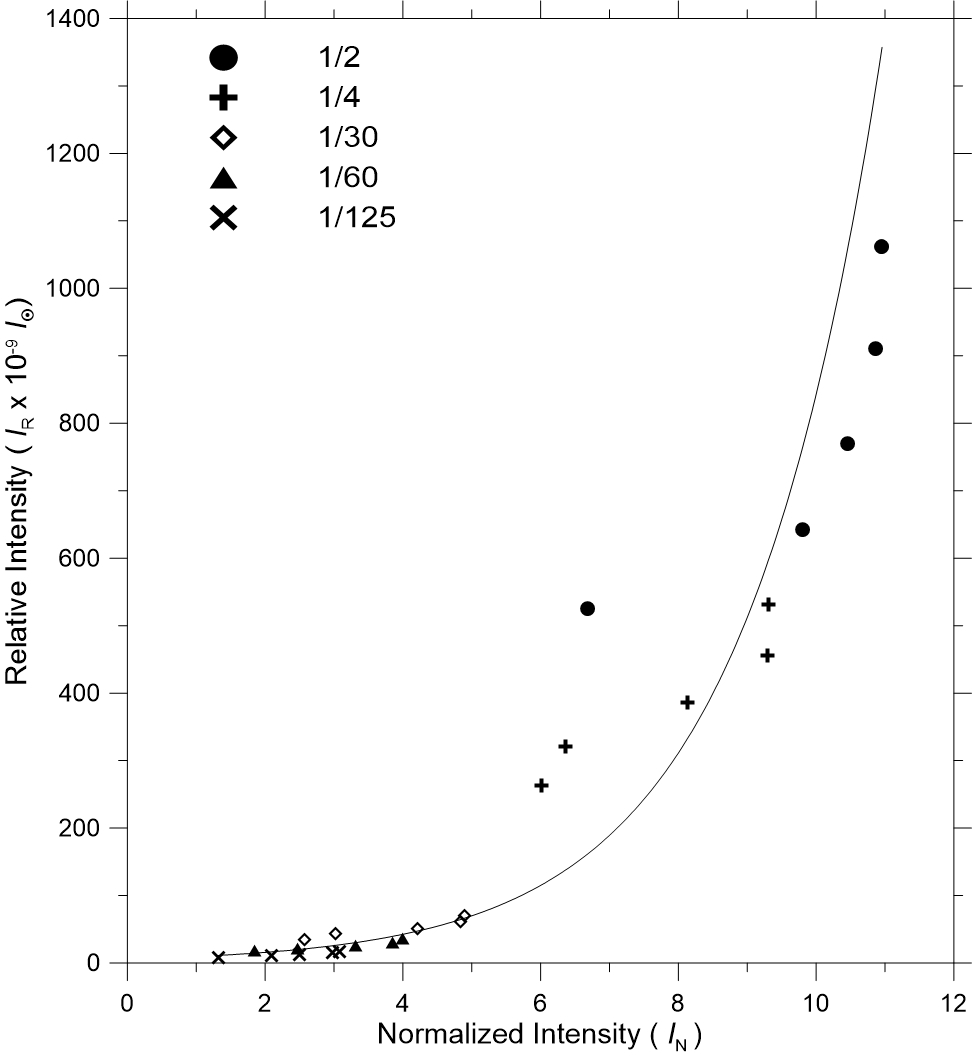}
	\caption{Graph drawn between normalized and relative intensities of TSE2006. Every exposure time is expressed with a different symbol, and the solid line represents the fit curve.}
\end{figure}
\newpage
\noindent The obtained fit curve is usually expressed in the form
\begin{equation} \label{eq:4}
	Y = A \hspace*{0.1cm} \mn{e}^{B \hspace*{0.05cm} X} \, .
\end{equation}
This is an exponential function notation and in this method, this curve is named the \qq{intensity calibration function} (ICF) for the used photographic material. The shape of the fit curve (in Figure 4) shows that this curve is partly similar to the curve described by \cite{HD1890} for density-log exposure. From Table 2 and Figure 4, this function for TSE2006 is obtained as
\begin{equation} \label{eq:5}
	I_\mn{R} = 5.7969 \times 10^{-9} \hspace*{0.1cm} \mn{e}^{0.4979 \hspace*{0.05cm} I_\mn{N}} \, ,
\end{equation}
with a correlation coefficient of $R^\mn{2}=0.92$. Based on Equation 2, the observed intensity is mainly a function of exposure time and the area of the telescope aperture, but the increase in the relative intensity with exposure is not linear, rather it is almost exponential (Figure 4). Because the light sensitivity of these films depends on the photographic materials that are used, their sensitivity is not linear, but almost logarithmic \citep{MCE1942}. The response of this material to light is rapid, and this shows itself as an exponential increase in Figure 4.

\section{Calculation of Combined Intensity}\label{sec:3}
Sufficient explanatory information on combining intensities is rarely found in the literature. This part of our study has a particular importance from this perspective. As explained previously, several photographs were taken with different exposures. These images should be combined reversely to accurately reveal the coronal intensity. The algorithm that combines intensities is mainly constructed by analyzing the equation of total coronal brightness in Stokes parameters. Here, three different intensity values are combined by the equation \citep{BDE1966}
\begin{equation} \label{eq:6}
	I = \frac{2}{3} \, \left( I_0 + I_{60} + I_{120} \right)
\end{equation} 
where $I_0$, $I_{60}$ and $I_{120}$ are intensities measured at $0^{\circ}$, $60^{\circ}$ and $120^{\circ}$ polarization angles, respectively. There are three intensities to take into account. The sum is therefore divided by three, which represents their average value. The factor of 2 as a multiplier comes from the equation \citep{GD2003}
\begin{equation} \label{eq:7}
	\left< {I_\mn{u}} \right> = 2 \, \left< I_\mn{pol} \right> \, ,
\end{equation}
where $I_\mn{u}$ and $I_\mn{pol}$ are unpolarized and polarized light, respectively. This was obtained by integrating the Malus law over whole angles, and it indicates that the unpolarized light intensity is twice that of the polarized light. As a first approach, the combined intensity of images taken with different exposures was calculated by dividing the sum of each image intensity by the number of exposures. For example, five different exposures of 1/2, 1/4, 1/30, 1/60 and 1/125 second, and three polarization angles $0^{\circ}$, $60^{\circ}$, and $120^{\circ}$, were used at TSE2006. Accordingly, the combined intensity of the $0^{\circ}$ polarization angle was calculated by
\begin{equation} \label{eq:8}
	I_\mn{C} (0^{\circ}) = \frac{I_{1/2} + I_{1/4} + I_{1/30} + I_{1/60} + I_{1/125}}{5}
	=\frac{\Sigma I_\mn{exp}}{n_\mn{exp}} \, ,
\end{equation}
where $I_{1/2}$ is measured intensity value for exposure time 1/2 second, $I_{1/4}$ is for 1/4 second and so on. The last part of the equation is a mathematical notation. $\Sigma I_{\mn{exp}}$ is the sum of the intensities and $n_{\mn{exp}}$ is the number of exposures. When the intensity is in question, the exposure time should be taken into account. According to the definition of an exposure given by \cite{HD1890}, this is a product of light intensity and time. Therefore the intensity on an image is proportional to the exposure time, and any observed intensity ($I_\mn{obs}$) can be defined by
\begin{equation} \label{eq:9}
	I_\mn{obs} \approx I_\mn{src} \; t \, ,
\end{equation}
where $I_\mn{src}$ is the source intensity, and {\it t} is the exposure time. Hence, because the combined intensity has resulted from the sum of different exposed intensities, the total exposure time ($\Sigma t$) of the combined intensity must be the sum of the exposure times that were used. For example, for TSE2006, this total exposure time is calculated as
\begin{equation}  \nonumber
	\Sigma t = \frac{1}{2}+\frac{1}{4}+\frac{1}{30}+\frac{1}{60}+\frac{1}{125} = 0.808 \hspace{0.1cm} 
	\textrm{second}.
\end{equation}
As a result, the combined intensity (I$_\mn{C}$) for any polarization angle can be given by
\begin{equation} \label{eq:10}
	I_\mn{C} = \frac{1}{\Sigma t} \frac{\Sigma I_\mn{exp}}{n_\mn{exp}}.
\end{equation}

\section{Calculated Total Brightness Values of the Solar Eclipse in 2006}\label{sec:4}
The results of the total brightness obtained in TSE2006 are presented as an example or application for the methods described in Sections 2.1 and 3. Several eclipse images are shown in Figure 5. The intensities of all eclipse images taken at different exposure times were combined using Equation 10 separately for each polarization angle. The composite images (or intensities) of the three polarization angles are shown in Figure 6.

The total coronal brightness values of TSE2006 were calculated using Equation 6, and the results are shown in Figure 7 with selected lines of equal intensities. After all the radial directions at polar angles (PA) were defined with respect to the North Pole of the Sun in the range from $0^{\circ}$ to $350^{\circ}$ with $10^{\circ}$ steps, the distance of every selected isophote was calculated separately through all radial directions. The average distances of all isophotes were obtained by reducing them into one quadrant ranging from $0^{\circ}$ to $90^{\circ}$ in order to compare them with coronal models and several studies of other authors. The calculated values are listed in Table 3 with selected isophote values.  

\begin{figure}[htbp]\label{fig:5}
	\centerline{\includegraphics[width=0.96\textwidth]{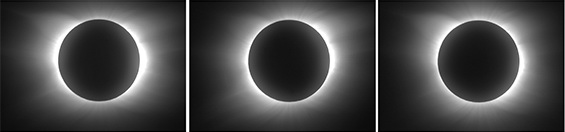}}
	\vspace{-5mm} \hspace{1mm} \color{white} \large 1/4 \vspace{0.5mm}
	
	\centerline{\includegraphics[width=0.96\textwidth]{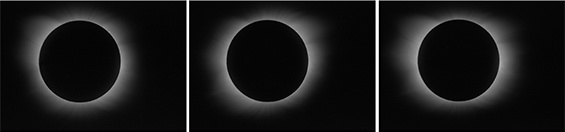}}
	\vspace{-5mm} \hspace{1mm} \color{white} \large 1/30 \vspace{0.5mm}
	
	\centerline{\includegraphics[width=0.96\textwidth]{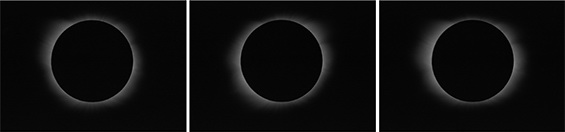}}
	\vspace{-5mm} \hspace{1mm} \color{white} \large 1/125 \\
	\hspace*{18mm} \color{black} {\normalsize $0^{\circ}$ \hspace{33mm} $60^{\circ}$ \hspace{33mm} $120^{\circ}$}
	\normalcolor 
	\caption{Several eclipse images taken at TSE2006. The numbers on the left are exposure times in seconds, and the numbers at the bottom are the polarization angles of the polarizer.}
\end{figure}

\begin{figure}[htbp]\label{fig:6}
	\centering	\includegraphics[width=0.92\textwidth]{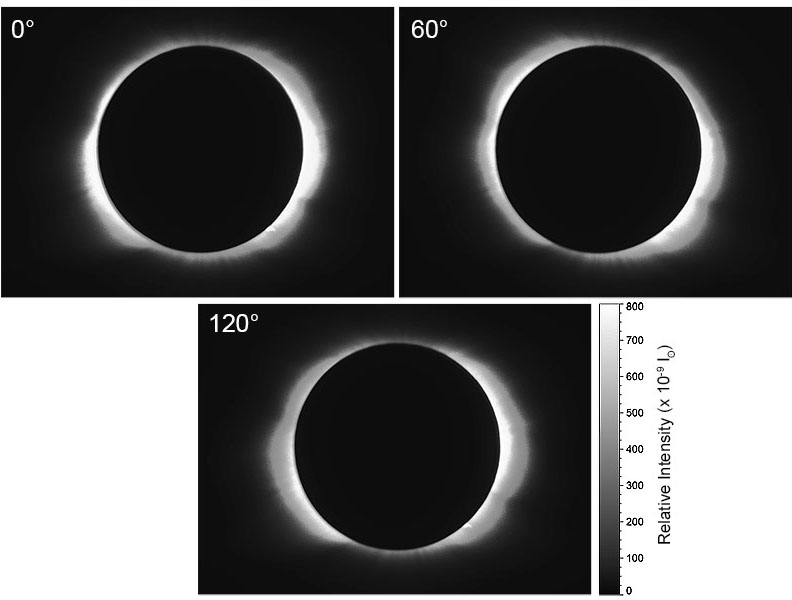}
	\caption{Composite images or combined intensity values of three polarization angles ($0^{\circ}$, $60^{\circ}$, and  $120^{\circ}$) for TSE2006.}
\end{figure}

\begin{figure}[h!]\label{fig:7}
	\centering \includegraphics[width=0.95\textwidth]{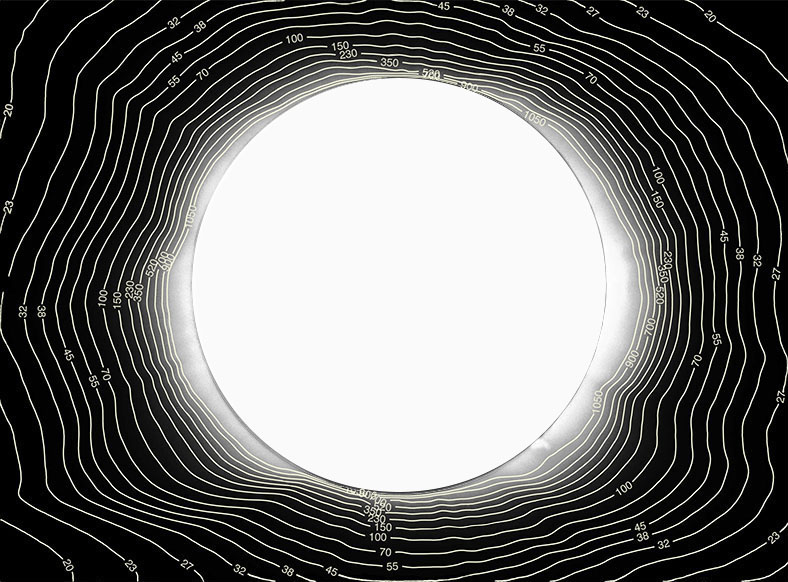}
	\caption{Image of total coronal brightness with isophotes for TSE2006. Isophote intensity values are displayed in units of $10^{-9} \bar I_{\odot}$.}
\end{figure}

\setlength{\tabcolsep}{3.7pt} 
\renewcommand{\arraystretch}{1.1}%
\newcommand{\bb}{\hspace{0.01cm}}
\begin{table}[h!] \label{tab:3}%
	%  \centering
	\caption{Average distances of the isophotes selected according to the radial direction of the polar angles.}
	\begin{tabular}{rccccccccccc}
		Iso. Int. &       & \multicolumn{10}{c}{Polar Angles} \\
		\cline{3-12}    
		$10^{-9} \bar I_{\odot}$ & Log {\textit I} & 0 & 10 & 20 & 30 & 40 & 50 & 60 & 70 & 80 & 90 \\
		\cline{1-12} 
		1050 \bb & -5.979 & 1.051 & 1.053 & 1.058 & 1.078 & 1.111 & 1.133 & 1.138 & 1.151 & 1.172 & 1.187 \\
		 900 \bb & -6.046 & 1.061 & 1.062 & 1.072 & 1.099 & 1.145 & 1.164 & 1.166 & 1.177 & 1.198 & 1.216 \\
		 700 \bb & -6.155 & 1.073 & 1.077 & 1.096 & 1.129 & 1.179 & 1.206 & 1.207 & 1.214 & 1.234 & 1.250 \\
		 520 \bb & -6.284 & 1.098 & 1.102 & 1.126 & 1.158 & 1.211 & 1.244 & 1.240 & 1.245 & 1.269 & 1.282 \\
		 350 \bb & -6.456 & 1.132 & 1.135 & 1.160 & 1.190 & 1.247 & 1.283 & 1.275 & 1.277 & 1.305 & 1.319 \\
		 230 \bb & -6.638 & 1.168 & 1.170 & 1.195 & 1.224 & 1.284 & 1.325 & 1.315 & 1.311 & 1.344 & 1.359 \\
		 150 \bb & -6.824 & 1.212 & 1.213 & 1.236 & 1.268 & 1.334 & 1.379 & 1.371 & 1.356 & 1.393 & 1.408 \\
		 100 \bb & -7.000 & 1.265 & 1.269 & 1.288 & 1.326 & 1.391 & 1.449 & 1.445 & 1.417 & 1.453 & 1.469 \\
		  70 \bb & -7.155 & 1.331 & 1.338 & 1.352 & 1.390 & 1.455 & 1.527 & 1.532 & 1.493 & 1.526 & 1.542 \\
		  55 \bb & -7.260 & 1.393 & 1.400 & 1.412 & 1.445 & 1.507 & 1.591 & 1.610 & 1.563 & 1.595 & 1.606 \\
		  45 \bb & -7.347 & 1.456 & 1.464 & 1.474 & 1.501 & 1.560 & 1.657 & 1.690 & 1.636 & 1.666 & 1.672 \\
		  38 \bb & -7.420 &       &       & 1.534 & 1.564 & 1.617 & 1.726 & 1.769 & 1.709 & 1.737 & 1.738 \\
		  32 \bb & -7.495 &       &       &       & 1.632 & 1.693 & 1.808 & 1.862 & 1.797 & 1.827 & 1.822 \\
		  27 \bb & -7.569 &       &       &       &       & 1.797 & 1.905 & 1.985 & 1.912 & 1.940 & 1.933 \\
		\cline{1-12}
	\end{tabular}
	\vspace{1mm}
\end{table}%
\begin{figure}[h!]\label{fig:8}
	\hspace{1.5cm}
	\includegraphics[width=0.6\textwidth]{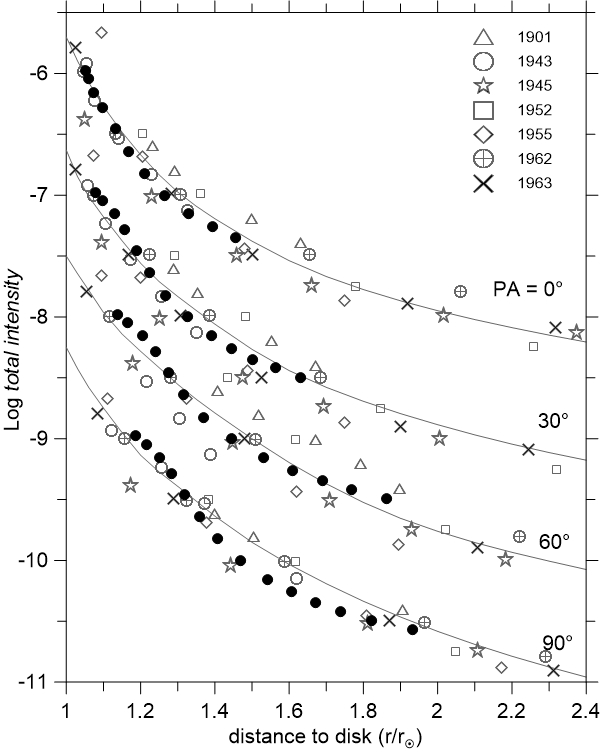}
	\caption{Total brightness values for four polar angles shown with black filled circles compared to the coronal model values of \cite{SK1970} (solid line) and observational values of seven different eclipses.}
\end{figure}

Total brightness values in the polar angles $0^{\circ}$, $30^{\circ}$, $60^{\circ}$ and $90^{\circ}$ are shown with black circles in Figure 8 compared to the minimum type coronal model (solid line) of \cite{SK1970} and seven eclipse observations made by various authors \citep{YRK1911,SK1948,RJM1951,SK1956,KH1958,SH1964,WM1964}. Each eclipse observation value is shown with a different symbol. The graph shows that the results of TSE2006 are in good agreement with both model and other observational values. 

\newpage
\section{Discussion}\label{sec:5}
An eclipse observation without polarization does not provide any information about the intensity of the K corona. White-light polarization observations are necessary to obtain the intensity of the K corona during totality, because the intensity of the K corona can only be calculated using Stokes parameters. The total intensity (polarized and unpolarized), polarization angles and polarization degree of the solar corona were calculated using the intensities of three different polarization angles (0$^{\circ}$, 60$^{\circ}$, and 120$^{\circ}$) \citep{BDE1966,GD2003}. After the total intensity was obtained, the contributions of unpolarized light, the sky, the instrumentations, and the F corona had be subtracted in order to obtain the intensity of the K corona.

In these studies calibrating the intensity is challenging and requires careful operations. The lack of information on this subject in the literature renders adds to this challenge. One of the aims of this study was to provide a straightforward approach for calculating the intensities of eclipse images and for obtaining the total brightness of the solar corona. The total brightness was obtained using another method, which combines the intensity values of different polarization angles. Another aim was generating an intensity calibration function that uses observational values. The developed calibration method is a practical method: only normalized and relative intensities are calculated. These parameters depend only on observational devices such as diaphragms, telescopes, and filters, and on parameters such as exposure time and the area of devices that the light passes through. This method was developed for all types of detectors.

Figure 4 shows that short-exposure values are located in the lower-left part of the curve and long-exposure values in the upper-right part. This means that the curve strictly depends on exposure times. The range of exposure times to be used therefore needs to be selected carefully to construct this calibration curve properly. The range also needs to be wide enough to show the calibration curve's sensitivity for both short- and long-exposure times. Short-exposure times must be selected as short as possible, whereas the long exposures should be kept as long as possible to give the opportunity of detecting the lower and upper intensity limit of the film sensitivity, respectively. 

One of the important parameters for eclipse observations can be obtained with the intensity calibration function, which gives the intensity contributions of the sky and instruments ($I_\mn{A+S}$) relative to the total measured intensity. As explained in Section 2.1, all intensities used in the calibration curve are normalized values, so that they are observational values. As a result of this the formed intensity calibration function is also observational. The lowest normalized intensity value in the calibration function gives the lowest relative intensity value indicating the far area from the solar disk, because the number 1 in the normalized intensity represents the relative background intensity. These far areas in the calibration images can be thought of areas that are not affected by the disk light. As a result of this approach, this intensity can be accepted as the sky intensity. Moreover, as observations are made using telescopes and other devices, their contributions to the observed intensity also needs to be taken into account. Thus, the lowest relative intensity represents the \qq{intensity contributions of the sky and instrumentation}. In the TSE2006 observation, this $I_{A+S}$ value was calculated as $0.95 \times 10^{-8} \bar I_{\odot}$ using Equation 5. The value obtained in this way is greater than the values of
\begin{tabbing}
	\quad $\bullet$ \hspace*{0.1mm} $0.18 \times 10^{-8}$ \quad \= 05 Feb 1962 eclipse\quad \= \cite{OS1967},\\
	\quad $\bullet$ \hspace*{0.1mm} $0.21 \times 10^{-8}$ \> 12 Nov 1966 eclipse \> \cite{NDS1970},\\
	\quad $\bullet$ \hspace*{0.1mm} $0.27 \times 10^{-8}$ \> 16 Feb 1980 eclipse \> \cite{DJ1982}, and\\
	\quad $\bullet$ \hspace*{0.1mm} $0.50 \times 10^{-8}$ \> 11 Aug 1999 eclipse \> \cite{KK2005}.
\end{tabbing}
The reason might be the telescope size (8-inch reflector). This telescope can analyze the solar corona up to 1.4R$_{\odot}$. On the other hand, the same method was used for the solar eclipse of 11 August 1999, and the obtained $I_\mn{A+S}$ value $0.63 \times 10^{-8} \bar I_{\odot}$ is in good agreement with the $0.5 \times 10^{-8} \bar I_{\odot}$ value of \cite{KK2005}, who observed the same eclipse. Parameters obtained from the calibration images taken at the 1999 eclipse and a graph of the intensity calibration function are also given in Appendix A with explanations and tables.

Methods for producing the composite images or combined intensity values of different polarization angles may be commonly known, but not enough explanatory sources are available in the literature. The method given in here is very simple to understand and has an easy-to-use formulation in general. Figure 8 shows that the obtained total intensities for TSE2006 agree well with the values given in the literature. The same agreement should be checked for in other eclipse observations. In the coming years, these methods will be tested as much as possible on new data of future eclipse observations.

\begin{acks}
Thanks to Ba\c{s}ar Co\c{s}kuno\u{g}lu for his contributions for improving the language of the manuscript. Thanks especially to M. T\"{u}rker \"{O}zkan and Adnan \"{O}kten and all other staff in charge of the observations of the 1999 and 2006 solar eclipses. Thanks also to anonymous reviewer for their valuable suggestions and comments that improved the manuscript. This work was supported by the Istanbul University Scientific Research Projects Commission with the project numbers 24242, UP-16/160399, and 470/27122005.
\end{acks}

\medskip \noindent
{\footnotesize \textbf{Disclosure of Potential Conflicts of Interest} The author declares that he has no conflicts of interest.}

\appendix 
\section{Results of the Calculation for the Intensity Calibration Function Made for the Solar Eclipse on 11 August 1999}\label{app:A}
\begin{table}[b!]\label{tab:4}%
	\renewcommand{\arraystretch}{1.01}    
	\setlength{\tabcolsep}{8pt}
	\caption{Measured intensities of the apparent solar disks and their background values of the calibration images taken at the solar eclipse on 11 August 1999.}
	\vspace{0.005\textheight}
	\begin{tabular}{*{5}{c}|*{4}{c}}  \cline{1-9}
		& \multicolumn{4}{c|}{exposures / solar Disk Intensity} & \multicolumn{4}{c}{exposures / Background Intensity} \\
		\cline{2-9}    Dia.Rad. & 1/2 & 1/4 & 1/15 & 1/60 & 1/2 & 1/4 & 1/15 & 1/60 \\
		\cline{1-9}
		1.0 & 2363 &      &      &      & 1413 &      &      &  \\
		1.5 & 3266 & 2873 &      &      & 1425 & 1408 &      &  \\
		2.5 & 4466 & 4375 & 2435 &      & 1272 & 1417 & 1473 &  \\
		3.5 & 5716 & 5213 & 3224 & 1931 & 1414 & 1389 & 1359 & 1424 \\
		4.5 & 6660 & 5928 & 3724 & 2304 & 1529 & 1351 & 1332 & 1215 \\
		5.5 & 6984 & 6797 & 4429 & 2722 & 1517 & 1446 & 1368 & 1215 \\
		6.5 & 7635 & 7095 & 4925 & 3515 & 1564 & 1437 & 1403 & 1525 \\
		7.5 & 8243 & 7546 & 5131 & 3846 & 1717 & 1534 & 1429 & 1378 \\
		8.5 &      & 7931 & 5827 & 4302 &      & 1762 & 1564 & 1397 \\
		9.5 &      &      & 5795 & 4289 &      &      & 1385 & 1344 \\
		10.5 &      &      &      & 4619 &      &      &      & 1360 \\
		\cline{1-9}
	\end{tabular}%
\end{table}%
The solar eclipse 1999 was observed by the staff of the Astronomy and Space Science Department of Istanbul University at the Tokat-Turhal observation site. The telescope named T150, which and has a 150 cm focal length and 13 cm aperture size, was used for the observation. A total of 11 square diaphragms with edge lengths of 1, 1.5, 2.5, 3.5, 4.5, 5.5, 6.5, 7.5, 8.5, 9.5 and 10.5 cm were used in the intensity calibration observation. Four different exposure times of 1/2, 1/4, 1/15, and 1/60 second were selected. The measured intensities of the apparent solar disks and their backgrounds are listed in Table 4 for every exposure. The average background intensity is obtained as 1215 from the intensities of the shortest exposures. The normalized and relative intensities of the calibration images were calculated and the results are listed in Table 5 with exposure times and diaphragm openings. In this observation, the light transmission of the solar filter and polarizer was $f_\mn{int}=1.6 \times 10^{-5}$ and $f_\mn{pol}=0.3$, respectively.
\renewcommand{\arraystretch}{0.95}    
\setlength{\tabcolsep}{8pt} 
\newcommand{\cwrt}[1]{\multicolumn{1}{c}{#1}}
\begin{table}[t!]\label{tab:5}
	\caption{Calculated normalized and relative intensities of the calibration images in the eclipse 1999.}
	\vspace{0.005\textheight}
	\begin{tabular}{*{5}{c}|*{4}{r}} \cline{1-9}
		& \multicolumn{4}{c|}{Normalized Intensity $I_\mn{N}$} & \multicolumn{4}{c}{Relative Intensity $I_\mn{R}$ ($\times 10^{-9} \bar I_{\odot} $)} \\
		\cline{2-9} 
		Dia.Rad. & 1/2 & 1/4 & 1/15 & 1/60 & 1/2 & 1/4 & 1/15 & 1/60 \\
		\cline{1-9}
		 1.0 & 1.95 &      &      &      &   18 &     &     &  \\
		 1.5 & 2.69 & 2.36 &      &      &   41 &  20 &     &  \\
		 2.5 & 3.68 & 3.60 & 2.01 &      &  113 &  56 &  15 &  \\
		 3.5 & 4.71 & 4.29 & 2.65 & 1.59 &  221 & 111 &  29 &  7 \\
		 4.5 & 5.48 & 4.88 & 3.07 & 1.89 &  366 & 183 &  49 & 12 \\
		 5.5 & 5.75 & 5.59 & 3.65 & 2.24 &  547 & 273 &  73 & 18 \\
		 6.5 & 6.29 & 5.84 & 4.05 & 2.89 &  764 & 382 & 102 & 25 \\
		 7.5 & 6.79 & 6.21 & 4.22 & 3.16 & 1017 & 509 & 136 & 34 \\
		 8.5 &      & 6.53 & 4.79 & 3.54 &      & 653 & 174 & 44 \\
		 9.5 &      &      & 4.77 & 3.53 &      &     & 218 & 54 \\
		10.5 &      &      &      & 3.80 &      &     &     & 66 \\
		\cline{1-9}
	\end{tabular}%   
\end{table}%
\begin{figure}[b!]\label{fig:9}
	\centering
	\includegraphics[width=0.57\textwidth]{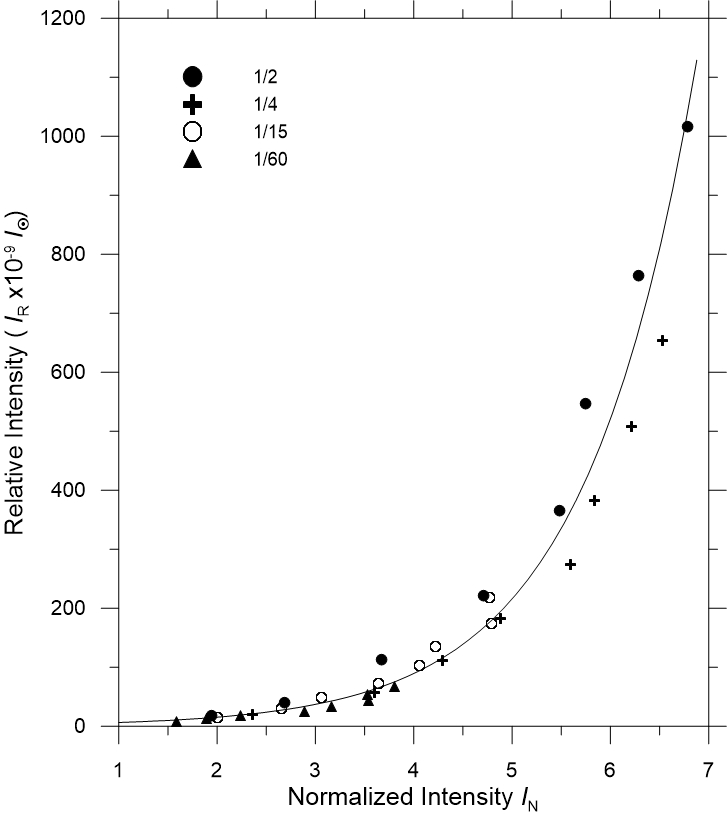}
	\caption{Graph for the normalized and relative intensities of the eclipse in 1999. The solid line represents the fitted curve, and the symbols represent different exposure times.}
\end{figure}A graph drawn between normalized and relative intensities is shown in Figure 9. The obtained intensity calibration function from this graph is given by
\begin{equation} \label{eq:12}
I_R = 2.6284 \times 10^{-9} \hspace*{0.1cm} \mn{e}^{0.8816 \hspace*{0.05cm} I_N} \, ,
\end{equation} 
where the correlation coefficient is $R^\mn{2}=0.97$. The sky and instrumental contributions value is calculated as $0.63 \times 10^{-8} \bar I_{\odot}$ for this observation.

\end{article} 
\end{document}